\documentclass[runningheads]{llncs}

\usepackage{graphicx}
\usepackage{amsfonts}
\usepackage{amsmath}
\usepackage{amssymb}
\usepackage{enumerate}
\usepackage{bm}
\usepackage{color}
\usepackage{wrapfig}

\usepackage{hyperref}

\begin{document}

\title{A machine learning based software pipeline to pick the variable ordering for algorithms with polynomial inputs}

\titlerunning{A pipeline to pick the variable ordering for algorithms with polynomial input}

\author{Dorian Florescu \and Matthew England}
\authorrunning{D. Florescu and M. England}

\institute{Faculty of Engineering, Environment and Computing, \\Coventry University, Coventry, CV1 5FB, UK
\email{\\ \{Dorian.Florescu, Matthew.England\}@coventry.ac.uk}}

\maketitle

\begin{abstract}
We are interested in the application of Machine Learning (ML) technology to improve mathematical software.  It may seem that the probabilistic nature of ML tools would invalidate the exact results prized by such software, however, the algorithms which underpin the software often come with a range of choices which are good candidates for ML application.  We refer to choices which have no effect on the mathematical correctness of the software, but do impact its performance.  

In the past we experimented with one such choice:  the variable ordering to use when building a Cylindrical Algebraic Decomposition (CAD).  We used the Python library Scikit-Learn (\texttt{sklearn}) to experiment with different ML models, and developed new techniques for feature generation and hyper-parameter selection.

These techniques could easily be adapted for making decisions other than our immediate application of CAD variable ordering.  Hence in this paper we present a software pipeline to use \texttt{sklearn} to pick the variable ordering for an algorithm that acts on a polynomial system.  The code described is freely available online.



\end{abstract}

\section{Introduction and context}
\label{SEC:Intro}

Mathematical Software, i.e. tools for effectively computing mathematical objects, is a broad discipline: the objects in question may be expressions such as polynomials or logical formulae, algebraic structures such as groups, or even mathematical theorems and their proofs.  In recent years there have been examples of software that acts on such objects being improved through the use of artificial intellegence techniques.  For example, \cite{KUV15} uses a Monte-Carlo tree search to find the representation of polynomials that are most efficient to evaluate; \cite{LGPC16b} uses a machine learnt branching heuristic in a SAT-solver for formulae in Boolean logic; \cite{GHK20} uses pattern matching to determine whether a pair of elements from a specified group are conjugate; and \cite{ACEISU16} uses deep neural networks for premise selection in an automated theorem proving.  See the survey article \cite{England2018} in the proceedings of ICMS 2018 for more examples.

Machine Learning (ML), that is statistical techniques to give computer systems the ability to \emph{learn} rules from data, may seem unsuitable for use in mathematical software since ML tools can only offer probabilistic guidance, when such software prizes exactness. However, none of the examples above risked the correctness of the end-result in their software.  They all used ML techniques to make non-critical choices or guide searches: the decisions of the ML carried no risk to correctness, but did offer substantial increases in computational efficiency.
All mathematical software, no matter the mathematical domain, will likely involve such choices, and our thesis is that in many cases an ML technique could make a better choice than a human user, so-called magic constants \cite{Carette2004}, or a traditional human-designed heuristic. 

\subsection*{Contribution and outline}

In Section \ref{SEC:OurWork} we briefly survey our recent work applying ML to improve an algorithm in a computer algebra system which acts on sets of polynomials.  We describe how we proposed a more appropriate definition of model accuracy and used this to improve the selection of hyper-parameters for ML models; and a new technique for identifying features of the input polynomials suitable for ML.   

These advances can be applied beyond our immediate application:  the feature identification to any situation where the input is a set of polynomials, and the hyper-parameter selection to any situation where we are seeking to take a choice that minimises a computation time.  Hence we saw value in packaging our techniques into a software pipeline so that they may be used more widely. Here, by pipeline we refer to a succession of computing tasks that can be run as one task. The software is freely available as a Zenodo repository here: \url{https://doi.org/10.5281/zenodo.3731703}

We describe the software pipeline and its functionality in Section \ref{SEC:Pipeline}.  Then in Section \ref{SEC:NewData} we describe its application on a dataset we had not previously studied.

\section{Brief survey of our recent work}
\label{SEC:OurWork}

Our recent work has been using ML to select the variable ordering to use for calculating a cylindrical algebraic decomposition relative to a set of polynomials.

\subsection{Cylindrical algebraic decomposition}
\label{SUBSEC:CAD}

A \emph{Cylindrical Algebraic Decomposition} (CAD) is a \emph{decomposition} of ordered $\mathbb{R}^n$ space into cells arranged \emph{cylindrically}, meaning the projections of cells all lie within cylinders over a CAD of a lower dimensional space.  All these cells are (semi)-algebraic meaning each can be described with a finite sequence of polynomial constraints.  A CAD is produced for either a set of polynomials, or a logical formula whose atoms are polynomial constraints.  It may be used to analyse these objects by finding a finite sample of points to query and thus understand the behaviour over all $\mathbb{R}^n$.  The most important application of CAD is to perform Quantifier Elimination (QE) over the reals.  I.e. given a quantified formula, a CAD may be used to find an equivalent quantifier free formula\footnote{E.g. QE would transform $\exists x, ax^2 + b x + c = 0 \land a \neq 0$ into the equivalent $b^2 - 4ac \geq 0$.\label{fn1}}.  

CAD was introduced in 1975 \cite{Collins1975} and is still an active area of research.  The collection \cite{CJ98} summarises the work up to the mid-90s while the background section of \cite{EBD20}, for example, includes a summary of progress since.  QE has numerous applications in science \cite{BDEEGGHKRSW20}, engineering \cite{Sturm2006}, and even the social sciences \cite{MDE18}. 

CAD requires an ordering of the variables.  QE imposes that the ordering match the quantification of variables, but variables in blocks of the same quantifier and the free variables can be swapped\footnote{In Footnote \ref{fn1} we must decompose $(x,a,b,c)$-space with $x$ last, but the other variables can be in any order.  Using $a \prec b \prec c$ requires 27 cells but $c \prec b \prec a$ requires 115\label{fn2}.}.  The ordering can have a great effect on the time / memory use of CAD, the number of cells, and even the underlying complexity \cite{BD07}.  Human designed heuristics have been developed to make the choice \cite{DSS04}, \cite{Brown2004}, \cite{BDEW13}, \cite{EBDW14} and are used in most implementations.  

The first application of ML to the problem was in 2014 when a support vector machine was trained to choose which of these heuristics to follow \cite{HEWDPB14}, \cite{HEWBDP19}.  The machine learned choice did significantly better than any one heuristic overall.

\subsection{Recent work on ML for CAD variable ordering}


The present authors revisited these experiments in \cite{EF19} but this time using ML to predict the ordering directly (because there were many problems where none of the human-made heuristics made good choices and although the number of orderings increases exponentially, the current scope of CAD application means this is not restrictive).  We also explored a more diverse selection of ML methods available in the Python library \texttt{scikit-learn} (\texttt{sklearn}) \cite{SciKitLearn2011}.  All the models tested outperformed the human made heuristics.  


The ML models learn not from the polynomials directly, but from features: properties which evaluate to a floating point number for a specific polynomial set. In \cite{HEWDPB14} and \cite{EF19} only a handful of features were used (measures of degree and frequency of occurrence for variables).  In \cite{FE19} we developed a new feature generation procedure which used combinations of basic functions (average, sign, maximum) evaluated on the degrees of the variables in either one polynomial or the whole system.  This allowed for substantially more features and improved the performance of all ML models.  The new features could be used for any ML application where the input is a set of polynomials.


The natural metric for judging a CAD variable ordering is the corresponding CAD runtime: in the work above models were trained to pick the ordering which minimises this for a given input.  However, this meant the training did not distinguish between any non-optimal ordering even though the difference between these could be huge.  This led us to a new definition of accuracy in \cite{FE20}: to picking an ordering which leads to a runtime within $x\%$ of the minimum possible.  

We then wrote a new version of the \texttt{sklearn} procedure which uses cross-validation to select model hyper-parameters to minimise the total CAD runtime of its choices, rather than maximise the number of times the minimal ordering is chosen.  This also improved the performance of all ML models in the experiments of \cite{FE20}.  The new definition and procedure are suitable for any any situation where we are seeking to take a choice that minimises a computation time.  

\section{Software pipeline}
\label{SEC:Pipeline}

The input to our pipeline is given by two distinct datasets used for training and testing, respectively. An individual entry in the data set is a set of polynomials that represent an input to a symbolic computation algorithms, in our case CAD.  The output is a corresponding sequence of variable ordering suggestions for each set of polynomials in the testing dataset. 

The pipeline is fully automated:  it generates and uses the CAD runtimes for each set of polynomials under each admissible variable ordering; uses the runtimes from the training dataset to select the hyper-parameters with cross-validation and tune the parameters of the model; and evaluates the performance of those classifiers (along with some other heuristics for the problem) for the sets of polynomials in the testing dataset.

We describe these key steps in the pipeline below. Each of the numbered stages can be individually marked for execution or not in a run of the pipeline (avoiding duplication of existing computation). The code for this pipeline, written all in Python, is freely available at: \url{https://doi.org/10.5281/zenodo.3731703}.

\subsection*{I. Generating a model using the training dataset}
	
\subsubsection*{(a) Measuring the CAD runtimes:}

The CAD routine is run for each set of polynomials in the training dataset. The runtimes for all possible variable orderings are stored in a different file for each set of polynomials. If the runtime exceeds a pre-defined timeout, the value of the timeout is stored instead.

\subsubsection*{(b) Polynomial data parsing:}

The training dataset is first converted to a format that is easier to process into features. For this purpose, we chose the format given by the \texttt{terms()} method from the \texttt{Poly} class located in the \texttt{sympy} package for symbolic computation in Python. 

Here, each monomial is defined by a tuple, containing another tuple with the degrees of each variable, and a value defining the monomial coefficient. The polynomials are then defined by lists of monomials given in this format, and a point in the training dataset consists of a list of polynomials. For example, one entry in the dataset is the set $\{235 x_1+42 x_2^2, 2 x_1^2 x_3-1\}$ which is represented as
				\begin{equation*}
				\left[\left[\left((1,0,0),235\right),\left((0,2,0),42\right)\right],\left[\left((2,0,1),2\right),\left((0,0,0),-1\right)\right]\right].
				\end{equation*}	
					
All the data points in the training dataset are then collected into a single file called \texttt{terms\_train.txt} after being placed into this format. Subsequently, the file \texttt{y\_train.txt} is created storing the index of the variable ordering with the minimum computing times for each set of polynomials, using the runtimes measured in Step I(a).

\subsubsection*{(c) Feature generation:}

Here each set of polynomials in the training dataset is processed into a fixed length sequence of floating point numbers, called features, which are the actual data used to train the ML models in \texttt{sklearn}. This is done with the following steps:

\begin{enumerate}[i.]				
\item {\bf Raw feature generation}\\
We systematically consider applying all meaningful combinations of the functions \texttt{average}, \texttt{sign}, \texttt{maximum}, and \texttt{sum} to polynomials with a given number of variables.  This generates a large set of feature descriptions as proposed in \cite{FE19}. The new format used to store the data described above allows for an easy evaluation of these features. An example of computing such features is given in Figure \ref{fig:featcomp}.
In \cite{FE19} we described how the method provides $1728$ possible features for polynomials constructed with three variables for example. This step generates the full set of feature descriptions, saved in a file called \texttt{features\_descriptions.txt}, and the corresponding values of the features on the training dataset, saved in a  file called \texttt{features\_train\_raw.txt}.

\begin{figure}[!ht]
\hfill
\begin{center}
\includegraphics[width=4.8in,trim={0cm 0cm 0cm 0cm},clip]{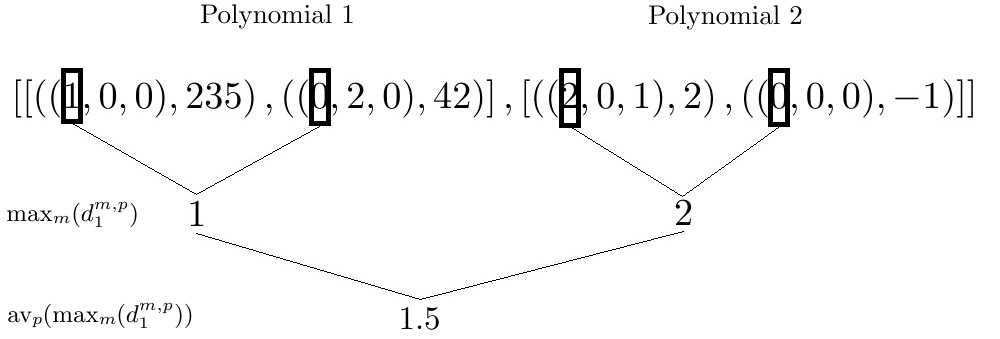}
\end{center}
\caption{Generating feature $\text{av}_p\left( \text{max}_m\left(d_1^{m,p}\right)\right)$ from data stored in the format of Section I(b).  Here $d_1^{m,p}$ denotes the degree of variable $x_1$ in polynomial number $p$ and monomial number $m$, and $\text{av}_p$ denotes the average function computed for all polynomials \cite{FE19}. \label{fig:featcomp}}
\end{figure}

\item {\bf Feature simplification}\\
After computing the numerical values of the features in Step I(c)i this step will remove those features that are constant or repetitive for the dataset in question, as described in \cite{FE19}. The descriptions of the remaining features are saved in a new file called \texttt{features\_descriptions\_final.txt}. \\
			
\item {\bf Final feature generation}\\
The final set of features is computed by evaluating the descriptions in \texttt{features\_descriptions\_final.txt} for the training dataset. Even though these were already evaluated in Step I(c)i we repeat the evaluation for the final set of feature descriptions. This is to allow the possibility of users entering alternative features manually and skipping steps i and ii.  As noted above, any of the named steps in the pipeline can be selected or skipped for execution in a given run. The final values of the features are saved in a new file called \texttt{features\_train.txt}.

\end{enumerate}

\subsubsection*{(d) Machine learning classifier training:}

\begin{enumerate}[i.]
\item{\bf Fitting the model hyperparameters by cross-validation}\\
The pipeline can apply four of the most commonly used deterministic ML models (see \cite{EF19} for details), using the implementations in \texttt{sklearn} \cite{SciKitLearn2011}.
\begin{itemize}
	\item The K-Nearest Neighbors (KNN) classifier
	\item The Multi-Layer Perceptron (MLP) classifier
	\item The Decision Tree (DT) classifier
	\item The Support Vector Machine (SVM) classifier
\end{itemize}
Of course, additional models in \texttt{sklearn} and its extensions could be included with relative ease.  
The pipeline can use two different methods for fitting the hyperparameters via a cross-validation procedure on the training set, as described in \cite{FE20}:
\begin{itemize}
	\item Standard cross-validation: maximizing the prediction accuracy (i.e. the number of times the model picks the optimum variable ordering).
	\item Time-based cross-validation: minimizing the CAD runtime (i.e. the time taken to compute CADs with the model's choices).
\end{itemize}
Both methods tune the hyperparameterswith cross-validation using the routine \texttt{RandomizedSearchCV} from the \texttt{sklearn} package in Python (the latter an adapted version we wrote).   The cross-validation results (i.e. choice of hyperparameters) are saved in a file \texttt{hyperpar\_D**\_**\_T**\_**.txt}, where \texttt{D**\_**} is the date and \texttt{T**\_**} denotes the time when the file was generated.

\item {\bf Fitting the parameters}\\
The parameters of each model are subsequently fitted using the standard sklearn algorithms for each chosen set of hyperparameters.  These are saved in a file called \texttt{par\_D**\_**\_T**\_**.txt}.
\end{enumerate}

\subsection*{II.  Predicting the CAD variable orderings using the testing dataset}

The models in Step I are then evaluated according to their choices of variable orderings for the sets of polynomials in the testing dataset. The steps below are listed without detailed description as they are performed similarly to Step I for the testing dataset.

\subsubsection*{(a) Polynomial data parsing:} 

The values generated are saved in a new file called \texttt{terms\_test.txt}.			

\subsubsection*{(b) Feature generation:} 

The final set of features is computed by evaluating the descriptions in Step I(b)ii for the testing dataset. These values are saved in a new file called \texttt{features\_test.txt}.		

\subsubsection*{(c) Predictions using ML:}
Predictions on the testing dataset are generated using the model computed in Step I(c). The model is run with the data in Step II(a)ii, and the predictions are stored in a file called \texttt{y\_D**\_**\_T**\_**\_test.txt}.

\subsubsection*{(d) Predictions using human-made heuristics:}

In our prior papers \cite{EF19}, \cite{FE19}, \cite{FE20} we compared the performance of the ML models with the human-designed heuristics in \cite{Brown2004} and \cite{DSS04}. For details on how these are applied see \cite{EF19}.  Their choices are saved in two files entitled \texttt{y\_brown\_test.txt} and \\ \texttt{y\_sotd\_test.txt}, respectively.

\subsubsection*{(e) Comparative results:}

Finally, in order to compare the performance of the proposed pipeline, we must  measure the actual CAD runtimes on the testing dataset. The results of the comparison is saved in a file with the template:\\ \texttt{comparative\_results\_D**\_**\_T**\_**.txt}.

\subsection*{Adapting the pipeline to other algorithms}

The pipeline above was developed for choosing the variable ordering for the CAD implementation in Maple's Regular Chains Library \cite{CMXY09}, \cite{CM16}. But it could be used to pick the variable ordering for other procedures which take sets of polynomials as input by changing the calls to CAD in Steps I(a) and II(e) to that of another implementation / algorithm.  Step II(d) would also have to be edited to call an appropriate competing heuristic.

\section{Application of pipeline to new dataset}
\label{SEC:NewData}

The pipeline described in the previous section makes it easy for us to repeat our past experiments (described in Section \ref{SEC:OurWork}) for a new dataset. All that is needed to do is replace the files storing the polynomials and run the pipeline. 

To demonstrate this we test the proposed pipeline on a new dataset of randomly generated polynomials.  We are not suggesting that it is appropriate to test classifiers on random data: we simply mean to demonstrate the ease with which the experiments in \cite{EF19}, \cite{FE19}, \cite{FE20} that originally took many man-hours can be repeated with just a single code execution.

The randomly generated parameters are: the degrees of the three variables in each polynomial term, the coefficient of each term, the number of terms in a polynomial and the number of polynomials in a set. The means and standard deviations of these parameters were extracted from the problems in the \texttt{nlsat} dataset\footnote{\url{https://cs.nyu.edu/~dejan/nonlinear/}}, which was used in our previous work \cite{EF19} so that the dataset is of a comparable scale.  
We generated $7500$ sets of random polynomials, where $5000$ were used for training, and the remaining $2500$ for testing. 

The results of the proposed processing pipeline, including the comparison with the existing human-made heuristics are given in Table \ref{tab:1}.  The prediction time is the time taken for the classifier or heuristic to make its predictions for the problems in the training set.  The total time adds to this the time for the actual CAD computations using the suggested orderings.  We do not report the training time of the ML as this is a cost paid only once in advance.  The virtual solvers are those which always make the best/worst choice for a problem (in zero prediction time) and are useful to show the range of possible outcomes.  We note that further details on our experimental methodology are given in \cite{EF19}, \cite{FE19}, \cite{FE20}.

As with the tests on the original dataset \cite{EF19}, \cite{FE19} the ML classifiers outperformed the human made heuristics, but for this dataset the difference compared to the Brown heuristic was marginal.  We used a lower CAD timeout which may benefit the Brown heuristic as past analysis shows that when it makes sub-optimal choices these tend to much worse.  We also note that the relative performance of the Brown heuristic fell significantly when used on problems with more than three variables in \cite{FE20}.  The results for the sotd heuristic are bad because it had a particularly long prediction time on this random dataset.  We note that there is scope to parallelize sotd which may make it more competitive.

\begin{table}[t]
	\caption{The comparative performance of DT, KNN, MLP, SVM, the Brown and sotd heuristics on the testing data for our randomly generated dataset. A random prediction, and the virtual best (VB) and virtual worst (VW) predictions are also included.}\label{tab:1}
	\begin{center}
		\begin{tabular}{|c|c|c|c|c|c|c|c|c|c|c|}
			\hline
			& DT & KNN & MLP & SVM & \texttt{Brown} & \texttt{sotd} & rand & VB & VW \\
			\hline	\textbf{Prediction time (s)} & $ 4.8\cdot e^{-4} $ & $ 0.68 $ & $ 2.8\cdot e^{-4} $ & $ 0.99 $ & $ 53.01 $ & $15\,819$  &  &  &  \\
			\hline	\textbf{Total time (s)} & $ 6\,548 $ & $ 6\,610 $ & $ 6\,548 $ & $ 6\,565 $ & $ 6\,614 $ &  $ 22\,313 $ & $ 16\,479 $ & $5\,610$ & $25\,461$ \\				
			\hline
			\end{tabular}
	\end{center}
\end{table}

\section{Conclusions}

We presented our software pipeline for training and testing ML classifiers that select the variable ordering to use for CAD, and described the results of an experiment applying it to a new dataset.    

The purpose of the experiment in Section \ref{SEC:NewData} was to demonstrate that the pipeline can easily train classifiers that are competitive on a new dataset with almost no additional human effort, at least for a dataset of a similar scale (we note that the code is designed to work on higher degree polynomials but has only been testes on datasets of 3 and 4 variables so far).  The pipeline makes it possible for a user to easily tuning the CAD variable ordering choice classifiers to their particular application area.  

\newpage

Further, with only a little modification, as noted at the end of Section \ref{SEC:Pipeline}, the pipeline could be used to select the variable ordering for alternative algorithms that act on sets of polynomials and require a variable ordering.  We thus expect the pipeline to be a useful basis for future research and plan to experiment with its use on such alternative algorithms in the near future.

\subsubsection*{Acknowledgements}  

This work is funded by EPSRC Project EP/R019622/1: \emph{Embedding Machine Learning within Quantifier Elimination Procedures}.  We thank the anonymous referees for their comments.

%
%
%
%

\bibliographystyle{splncs04}

\end{document}